\documentclass[journal]{IEEEtran}

\IEEEoverridecommandlockouts                             
\usepackage{rotfloat}                                                   \usepackage{multirow} 
\usepackage{rotating}
\usepackage{graphics} 
\usepackage{epsfig} 
\usepackage{mathptmx} 
\usepackage{times}
\usepackage{amsmath}
\usepackage{amssymb}  
\usepackage{float}
\usepackage[table]{xcolor}

\title{A Novel Hierarchical Intrusion Detection System based on Decision Tree and Rules-based Models}

\author{Ahmed Ahmim$^{1}$, Leandros Maglaras$^{2}$, Mohamed Amine Ferrag$^{3}$, Makhlouf Derdour$^{1}$,  Helge Janicke$^{2}$
\thanks{$^{1}$A. Ahmim and M. Derdour are with Departement of  Mathematics and Computer Science,
        University of Larbi Tebessi, Tebessa, Algeria
        {\tt\small ahmed.ahmim@univ-tebessa.dz, makhlouf.derdour@univ-tebessa.dz}}
\thanks{$^{2}$L. A. Maglaras and Helge Janicke are with School of Computer Science and Informatics, De Montfort University, Leicester, United Kingdom
        {\tt\small leandros.maglaras@dmu.ac.uk, heljanic@dmu.ac.uk}}
        \thanks{$^{3}$M. A. Ferrag is with Department of Computer Science, Guelma University, B.P. 401, 24000, Algeria
        {\tt\small ferrag.mohamedamine@univ-guelma.dz}}
               }
\begin{document}

\maketitle
\thispagestyle{empty}
\pagestyle{empty}

\begin{abstract}
This paper proposes a novel intrusion detection system (IDS) that combines different classifier approaches which are based on decision tree and rules-based concepts, namely, REP Tree, JRip algorithm and Forest PA. Specifically, the first and second method take as inputs features of the data set, and classify the network traffic as Attack/Benign. The third classifier uses features of the initial data set in addition to the outputs of the first and the second classifier as inputs. The experimental results obtained by analyzing the proposed IDS using the CICIDS2017 dataset, 
attest their superiority in terms of accuracy, detection rate, false alarm rate and time overhead as compared to state of the art existing schemes.

\end{abstract}

\section{INTRODUCTION}

In recent years cyber attacks, especially those targeting systems that keep or process sensitive information are becoming more sophisticated. Critical National Infrastructures are main targets of cyber attacks, since essential information or services depend on their systems and their protection becomes a signiﬁcant issue that is concerning both organizations and nations \cite{maglaras2018cyber}. Attacks to such critical systems include penetrations to their network and installation of malicious tools or programs that can reveal sensitive data or alter the behavior of specific physical equipment. In order to tackle this growing trend academics and industry professionals are joining forces in an attempt to develop novel systems and mechanisms that can defend their systems. Along with other preventive security mechanisms, such as access control and authentication, intrusion detection systems (IDS) are deployed as a second line of defense. IDS based on some specific rules or patterns of normal behavior of the system can distinguish between normal and malicious actions \cite{alcaraz2015critical}. 

Many different taxonomies for IDSs have been proposed until now. Based on the classification model they use, IDSs can be classified as rule based, misuse detection and mixed systems.  IDSs can also be classified as or real time if they use contiguous monitoring of the system or periodic or off line if the detection happens in specific time instances or even off line using data that are collected and stored during a certain period of time.   Moreover, when talking about Industrial Control Systems (ICS) that have specific requirements and characteristics novel taxonomies were recently proposed. Authors in \cite{hu2018survey} proposed a classification of IDSs ICS that divides them in three new categories: protocol analysis-based, traffic mining-based, and control process analysis-based.

Countermeasures are taken accordingly to the information obtained regarding the detected attacks from the detection systems. The better classification of the type of the attack is provided, the more efficient countermeasures will be chosen and the less those will affect the proper operation of the system or network. Moreover, if we do not detect the exact type of attack the countermeasures can have more serious consequences than the attack itself in some cases. For this reason, our goal is to create an intrusion detection model that correctly classifies each type of attack. In addition, our model must provide a low false alarm rate and a high detection rate both for frequent and infrequent attacks while on the same require low computing in order to perform classiﬁcation. The latter characteristic is very important when IDSs are deployed in industrial control systems that operate critical infrastructures where correct and fast notification about cyber attacks is crucial \cite{cruz2016cybersecurity} 

The remainder of the paper is organized as follows. Section \ref{sec_related} discusses related work and places the research within that of the wider community. Section \ref{sec_model} introduces the key concepts and the overall architecture of the proposed system.  Section \ref{sec_exper} presents the experimentation setup, gives the simulation parameters and describes the evaluation of the method. Section \ref{sec_concl} concludes the paper.

\begin{table*}[h!]
\scriptsize
\label{tab1}
\centering
\caption{Related works on hybrid intrusion detection systems}
\begin{tabular}{|p{0.2in}|p{0.8in}|p{2.7in}|p{1.8in}|p{0.5in}|} \hline \hline
\textbf{Year} & \textbf{Paper} & \textbf{Machine learning and data mining methods} & \textbf{Cyber approach } & \textbf{Data used } \\ \hline 
2009 & Aydin et al. \cite{F1} & - Packet header anomaly detection\newline - Network traffic anomaly detection & Hybrid IDS, which combing anomaly-based IDSs & IDEVAL \\ \hline 
2010 & Wang et al. \cite{F2} & - Artificial neural networks\newline Fuzzy clustering & Hybrid IDS, which the fuzzy aggregation module is employed to aggregate the results & KDD CUP 1999 \\ \hline 
2011 & Govindarajan and Chandrasekaran \cite{F3} & - Multilayer perceptron neural network\newline - Radial basis function neural network & Neural based hybrid IDS & UNM Send-Mail Data \\ \hline 
2012 & Chunga and Wahidb \cite{F4} & - Intelligent dynamic swarm\newline - Simplified swarm optimization & Hybrid IDS & KDD CUP 1999 \\ \hline 
2013 & Elbasiony et al. \cite{F5} & - Random forests algorithm\newline - K-means clustering algorithm & Combining misuse and anomaly detection into a hybrid framework & KDD CUP 1999 \\ \hline 
2014 & Kim et al. \cite{F6} & - C4.5 decision tree algorithm\newline - Support vector machine model & Combining misuse and anomaly detection into a hybrid framework & NSL-KDD \\ \hline 
2015 & Lin et al. \cite{F7} & - k-Nearest Neighbor (k-NN) classifier & Combining cluster centers and nearest neighbors & KDD CUP 1999 \\ \hline 
2016 & Aslahi-Shahri et al. \cite{F8} & - Support vector machine\newline - Genetic algorithm & Hybrid IDS & KDD CUP 1999 \\ \hline 
2017 & Kevric et al. \cite{F9} & - Random tree\newline - C4.5 decision tree algorithm\newline - NBTree & Combining classifier model based on tree-based algorithms & NSL-KDD \\ \hline 
2017 & Al-Yaseen et al. \cite{F10} & - Support vector machine\newline - Extreme learning machine\newline - K-means clustering algorithm & Hybrid IDS & KDD CUP 1999 \\ \hline 
2018 & Ahmim et al. \cite{F11} & - Repeated Incremental Pruning to Produce Error, Reduction (RIPPER), RBF Network (RBFN), Ripple-down rule learner (Ridor), and Random Forests\newline - Naive Bayes (NB) & Combining probability predictions of a tree of classifiers & KDD CUP 1999 +\newline NSL-KDD\newline  \\ \hline 
2018 & Aljawarneh et al. \cite{F12} & J48, Meta Pagging, RandomTree, REPTree, AdaBoostM1, DecisionStump, and NaiveBayes & Hybrid IDS, which combing anomaly-based IDSs & NSL-KDD \\ \hline 
 & Our work  & - REP Tree\newline - JRip algorithm\newline - Random Forest & Hybrid IDS, which combining classifier model based on tree-based algorithms & CICIDS2017 \\ \hline \hline
 \end{tabular}
\begin{itemize}
 \item IDEVAL: MIT Lincoln Laboratories network traffic data
 \item KDD CUP 1999:  The data set is based on DARPA 1998 TCP/IP data and has basic features captured by pcap (with about 4 million records of normal and attack traffic)
 \item UNM Send-Mail Data: The data set is based on an immune system developed at the University of New Mexico
 \item NSL-KDD:  A modified version of KDD'99 data set, which it does not include redundant records in the train set.
 \item CICIDS2017: The dataset contains benign and the most up-to-date common attacks, which resembles the true real-world data (PCAPs) developed at Canadian Institute for Cybersecurity (University of New Brunswick) \cite{F13}
\end{itemize}
\end{table*}

\section{Relevant work}\label{sec_related}

Table \ref{tab1} lists the representative related works on hybrid intrusion detection systems, including, machine learning and data mining methods used, security issue they try to address along with the dadaset used to evaluate thier performance.

Aydin et al. \cite{F1} proposed a hybrid IDS by combining two approaches, namely, 1) packet header anomaly detection (PHAD), and 2) network traffic anomaly detection (NETAD) with the signature-based IDS Snort. Both PHAD and NETAD methods are anomaly-based IDSs. The ayd$\imath$n et al.'s system is tested on IDEVAL data, which shows that the number of attacks detected increases significantly using the proposed hybrid IDS as compared to signature-based systems. Authors in \cite{F8} presented an anomaly detection technique based on support vector machine (SVM) and genetic algorithm (GA), in order to improve the performance of classification for SVMs. The experimental results on the KDD CUP 1999 dataset show an outstanding true-positive value of 0.973 that comes with a 0.017 of false-positive value.

Wang et al. \cite{F2} proposed an intrusion detection approach, named FC-ANN, based on artificial neural networks (ANN) and fuzzy clustering. The FC-ANN approach uses three main modules, i.e., fuzzy clustering module, ANN module, and fuzzy aggregation module. The fuzzy clustering module is used to partition a given set of data into clusters. The ANN module is used to learn the pattern of every subset. The fuzzy aggregation module is used to aggregate different ANN's result and reduce any detection errors. The FC-ANN approach was tested on the KDD CUP 1999 dataset and was proven to be efficient against low-frequent attacks, i.e., R2L and U2R attacks.

Govindarajan and Chandrasekaran \cite{F3} proposed a neural-based hybrid IDS architecture using two methods, namely, 1) multilayer perceptron neural network (MLP), and 2) radial basis function neural network (RBF). The procedures of hybrid modeling using bagging classifiers are employed in order to increase robustness, accuracy, and better overall generalization. Moreover, the UNM Send-Mail Data is used in this study, which is based on an immune system developed at the University of New Mexico. 
The performance of the proposed IDS in terms of accuracy was 98.88\% and 94.31\% for normal and abnormal traffic respectively slightly better as compared to the single classifiers that compose it.


Chunga and Wahidb \cite{F4}  approach the problem of decision rules generation by employing intelligent dynamic swarm based rough set (IDS-RS) for feature selection and simplified swarm optimization with weighted local search (SSO--WLS) strategy for data classification. The study provides a full system solution for improving the searching process in SSO rule mining by weighing three predetermined constants. The experimental results on the KDD CUP 1999 dataset show that the proposed hybrid network intrusion detection system using intelligent dynamic swarm based rough set, shows a good overall performance with 93.3\% accuracy in average of 20 runs.

\begin{figure*}[h!]
\centering
\includegraphics[width=1\linewidth,height=9.5cm]{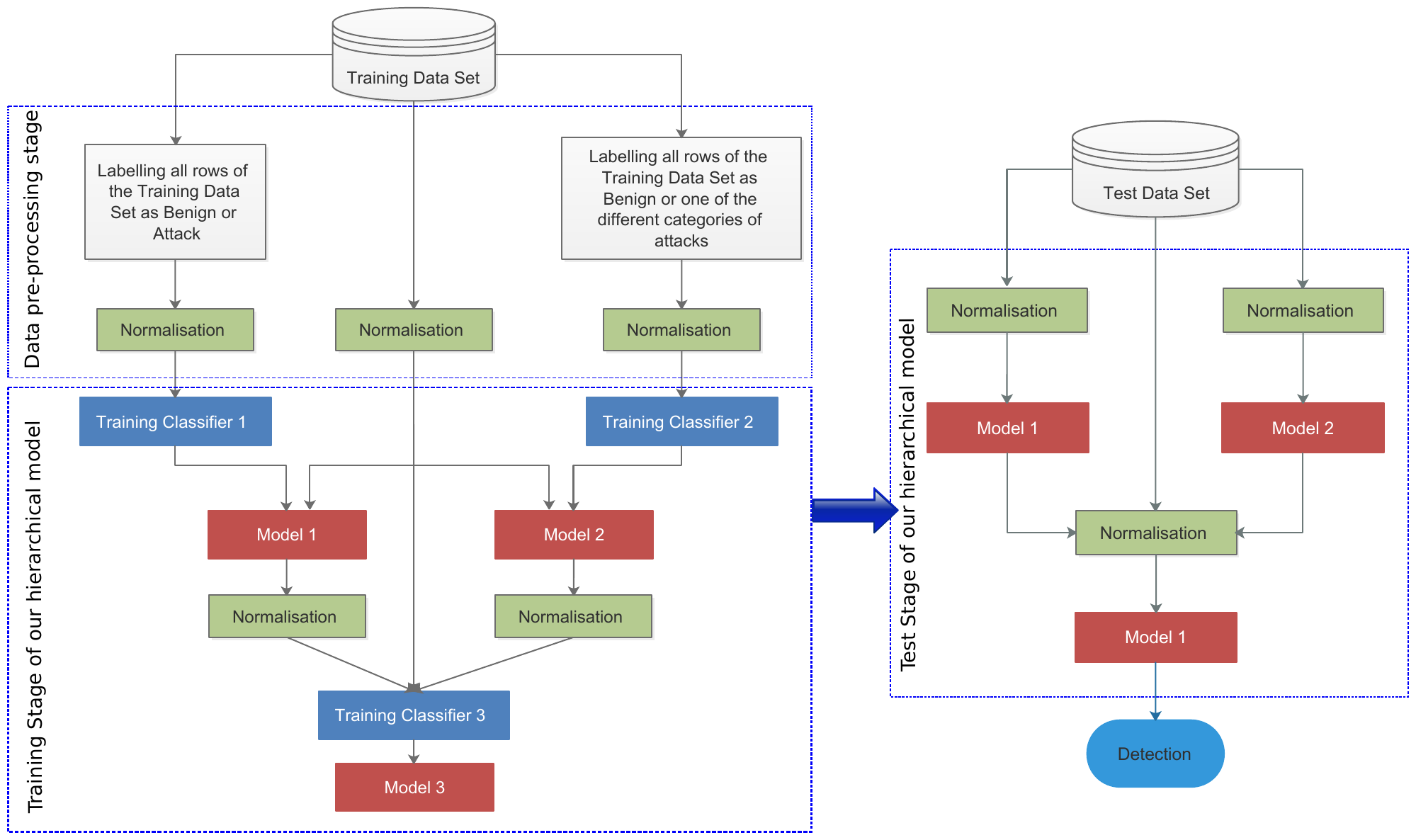}
	\caption{General structure of our proposed model.\label{F1}}
\end{figure*}

Elbasiony et al. \cite{F5} presented a combination of misuse and anomaly detection into a hybrid framework, which is based on two methods, namely, 1) random forests algorithm, and 2) K-means clustering algorithm. Specifically, this framework employs the random forests algorithm in misuse intrusion detection as well as the k-means clustering algorithm in anomaly detection. Due to correlated variables in random forests, this framework has low model interpretability and performance loss. Another study \cite{F6} integrates a misuse detection model and an anomaly detection model in a decomposition structure. This study uses the C4.5 decision tree algorithm and multiple one-class SVM models. The experimental results on the NSL-KDD dataset show that the proposed hybrid intrusion detection method is better than the conventional methods in terms of detection performance, training time, and testing time.
Lin et al. \cite{F7} proposed a feature representation approach, named CANN, which combining cluster centers and nearest neighbors. The CANN approach uses three steps, namely, 1)  Extraction of cluster centers and nearest neighbors, 2) Measurement and summing of the distance between all data, and 3) Classifier training and testing, which is based on the k-NN algorithm. The experimental results on the KDD CUP 1999  dataset show that CANN approach performs better than the k-NN and SVM classifiers.

Recently, a number of researchers have proposed the combination of classifiers in order to improve the overall performance under the NSL-KDD dataset \cite{F9,F10,F11,F12,F13}. In \cite{F9}, Kevric et al. proposed a combining classifier approach using tree algorithms, namely, random tree, C4.5 decision tree algorithm, and NBTree. In \cite{F10}, Al-Yaseen et al. proposed a hybrid IDS that uses support vector machine, extreme learning machine, and K-means clustering algorithm. The experimental results on the KDD CUP 1999 dataset show that the proposed hybrid network intrusion detection system can improve the overall performance and achieve an overall 95.75\% accuracy. 

In order to evaluate the performance of an IDS system, both KDD'99 and NSL-KDD datasets are commonly used. In \cite{F11}, Ahmim et al. proposed an IDS system, named HCPTC-IDS, which is based on combining probability predictions of a tree of classifiers. The HCPTC-IDS system is composed of two layers, namely, 1) the first layer, which is a tree of classifiers, and 2) The second layer, which is a final classifier that combines the different probability predictions of the first layer. The experiments on KDD'99 and NSL-KDD show that the HCPTC-IDS system is more precise than other recently proposed intrusion detection systems having accuracy equal to 96.27\% for KDD'99 and 89.75\% for NSL-KDD. 

Most of the relevant works use the KDD and NSL-KDD datasets, which are outdated and of very limited practical value for a modern IDS. Both benign and malicious network traffic has changed significantly since 1999 when these datasets were produced and the results obtained using them are of a limited value most of the times. 

In order to overcome shortcoming of previous proposed methods, like low detection of rare attack, miss-classification of attacks and time overhead we propose a novel Hybrid IDS, which combines different classifier models, namely, REP Tree, JRip algorithm and Random Forest. Besides, we use the CICIDS2017 dataset \cite{F13}, which we split in training and testing datasets, in order to evaluate their performances in detecting network intrusions and we compare it with other machine learning methods proposed by previous researchers, including, WISARD \cite{wisard}, ForestPA  \cite{ForestPA}, J48 Consolidated  \cite{j48Consolidated}, LIBSVM \cite{libsvm}, FURIA \cite{furia}, RandomForest, 
 REPTree, 
 MLP, 
 NaiveBayes, 
 Jrip 
 and J48. 

\section{Proposed model}\label{sec_model}
Based on the values of the features the classifier makes rules which are used in order to correctly classify the data set. In general, for the same data set,  if the number of classes increases, the sub-data set of each class decreases, leading to the decrease of the generalization capability  and the increase of classification errors and vice versa. The miss-classification is usually due to the classification of the attacks as normal behavior, or as another attack in the same or different category.
In order to minimize these classification errors and to increase the performance of intrusion detection mechanism, we propose the intrusion detection hierarchical model illustrated in figure \ref{F1}.

Our hierarchical model aims to detect the correct type  of each attack, provides a  low  false  alarm  rate  and  a  high  detection rate. It is composed of three classifiers. The first one uses the different features of the data set as inputs, in order to classify each row as benign or attack. The second one uses different features of the data set as inputs for  classifying each row as benign or one of the different categories of attacks. The third classifier uses all the features of the initial data set in addition to the outputs of the first and the second classifier as inputs, in order to classify each row of the data set as benign or a specific type of attack.

\subsection{Operation Mode}
The operation mode of our proposed model is composed of two steps: training  and testing.
\subsubsection{Training Step}
In this step, we train the three classifiers that compose our hierarchical model. We start by training the first and the second one and using the results from these two we train the third classifier.
Initially, the training Data set is labelled as benign and specific type of Attack. 
To train the first classifier, we modify the training data set, by labelling each row as "Attack" and "Benign". Then, we normalize the different features of the data set. After that, we perform  the training of this classifier and as result of this sub step, we get  model 1.
To train the second classifier, we modify the training data set, where use initial labelling of rows where each specific attack is identified. Then, we normalize the different features of the data set. After that, we perform the training of this classifier and as result of this sub step, we get  model 2. 

In order to train the third classifier, we modify the training data set, where we add two columns, the first one represents the classification results of model 1 for the rows of the training data set and the second one represents the classification results of model 2 for the rows of the training data set. Then, we normalize the different features of the data set. After that, we perform the training of this classifier and as result of this sub step, we get model 3. 

\subsubsection{Test Step}
As illustrated in Figure \ref{F1}, in order to test our hierarchical model, we process each row of the test data set by model 1 and model 2, then we add the outputs of model 1 and model 2 to the features of each row, and finally, we process the result rows by model 3 that classify it as Benign or a specific type of attack.

\section{Experimentation}\label{sec_exper}

In this section, we present in detail the Data Set used along with the Data pre-processing procedure. We also give the performance metrics used in our experiments. Moreover, we present the structure of our model. Finally, we provide a comparative study between our model and that of different classifiers. The experiments are done on a Windows 10 - 64 bits PC with 8 GB RAM and CPU Intel(R) I5 2.7 GHz. Weka Data Mining Tools and MySQL data base are used to implement our model.

\subsection{Data set and Data pre-processing} \label{data_preprocess}
In our experiments, we use  CICIDS 2017 \cite{F13}, which represents a data set that satisfy the eleven indispensable characteristics of a valid IDS dataset, namely Anonymity,  Attack Diversity, Complete Capture, Complete Interaction, Complete Network Configuration, Available Protocols, Complete Traffic, Feature Set, Metadata, Heterogeneity, and Labelling \cite{G16}.

CICIDS 2017 contains 2,830,743 rows devised on 8 files, each row having 79 features. Each row of CICIDS 2017 is labelled as Benign or one of fourteen type of attack.
Table \ref{T2} summarizes the distribution of different attack type and Benign rows.

In order to create a training and test subset,  we concatenate the 8 files in one same table that contains all benign and attacks rows. Then, we remove all rows that have the feature "Flow Packets/s" equal to 'Infinity' or 'NaN'.  After that, we remove the features that have a the same value for all rows, namely Bwd PSH Flags,Bwd URG Flags, Fwd Avg Bytes/Bulk, Fwd Avg Packets/Bulk, Fwd Avg Bulk/Rate, Bwd Avg Bytes/Bulk, Bwd Avg Packets/Bulk, Bwd Avg Bulk/Rate and Fwd Avg Bytes/Bulk.

After the elimination of these features, we extract the training and test subsets based on the distribution described in table \ref{T2}. In each subset we tried to include rows that contain all the attacks but the same row cannot appear in both subsets. 
For the training sub set, we select the first rows of each type. Then, For the test sub set, we select randomly the rows after the suppression of the training sub set rows.
\vspace{-1em}
\begin{table}[h!]
\scriptsize
\centering
\caption{Composition of Training and Test sub-sets}
\label{T2}
\begin{tabular}{|p{1.1cm}| p{1.3cm}|  p{0.9cm}| p{1.5cm}| p{0.8cm} |p{0.8cm}|}
\hline \hline
\multicolumn{2}{|c|}{Label}                                             & Total   & Total(-rows with lack info) & Training & Test \\ \hline
BENIGN                                     & BENIGN                     & 2273097 & 2271320                      & 20000    & 20000 \\ \hline
\multicolumn{1}{|c|}{\multirow{6}{*}{DOS}} & DDoS                       & 128027   & 128025       &      2700	& 3300            \\ \cline{2-6} 
\multicolumn{1}{|c|}{}                     & DoS slowloris              & 5796    & 5796                    &    1350 &	1650             \\ \cline{2-6} 
\multicolumn{1}{|c|}{}                     & DoS Slowhttptest           & 5499    & 5499                 &        2171 &	1169           \\ \cline{2-6} 
\multicolumn{1}{|c|}{}                     & DoS Hulk                   & 231073  & 230124             &          4500 &	5500           \\ \cline{2-6} 
\multicolumn{1}{|c|}{}                     & DoS GoldenEye              & 10293   & 10293                   &    1300 &	700           \\ \cline{2-6} 
\multicolumn{1}{|c|}{}                     & Heartbleed                 & 11      & 11                           & 5        & 5         \\ \hline
PortScan                                   & PortScan                   & 158930  & 158804                &      3808 &	4192      \\ \hline
Bot                                        & Bot                        & 1966    & 1956                         & 936 &	624        \\ \hline
\multirow{2}{*}{Brute-Force}               & FTP-Patator   & 7938    & 7935                         & 900 &	1100          \\ \cline{2-6} 
  & SSH-Patator  & 5897    & 5897   & 900 &	1100            \\ \hline
\multirow{3}{*}{Web Attack}                & Web Attack-Brute Force   & 1507    & 1507                         & 910 &	490      \\ \cline{2-6} 
   & Web Attack-XSS  & 652     & 652          & 480 &	160      \\ \cline{2-6} 
 & Web Attack-Sql Injection & 21      & 21   & 16 &	4      \\ \hline
Infiltration                               & Infiltration               & 36      & 36                           & 24 &	6         \\ \hline
\multicolumn{2}{|l|}{Total Attack}                                      & 471454  & 470365                       & 20000 &	20000     \\ \hline
\multicolumn{2}{|l|}{Total}                                             & 2830743 & 2827876                      & 40000 &	40000   \\ \hline \hline
\end{tabular}
\end{table}
\vspace{-2.4em}
\begin{table}[h!]
\scriptsize
\caption{Confusion matrix}
\label{TCM}
\begin{tabular}{cc|c|c|}
\cline{3-4}
\multicolumn{2}{c}{}                                                        & \multicolumn{2}{|c|}{Predicted class}                                \\ \cline{3-4} 
\multicolumn{1}{l}{}                                & \multicolumn{1}{l|}{} & \multicolumn{1}{l|}{Negative class} & \multicolumn{1}{l|}{Positive} \\ \hline
\multicolumn{1}{|c|}{\multirow{2}{*}{Actual class}} & Negative Class        & True negative (TN)                  & False positive (FP)           \\ \cline{2-4} 
\multicolumn{1}{|c|}{}                              & Positive  Class       & False negative (FN)                 & True positive (TP)            \\ \hline
\end{tabular}
\end{table}
\begin{table*}
\scriptsize
\centering
\caption{Performance of our model and other classifiers relative to the different attack type and Benign}
\label{T_SP}
\begin{tabular}{|p{1.5cm}|p{0.8cm}|p{0.9cm}|p{0.8cm}|p{1.3cm}|p{0.8cm}|p{0.7cm}|p{1.15cm}|p{0.8cm}|p{0.7cm}|p{1.1cm}|p{0.7cm}|p{0.7cm}|}

\hline
                                     & \textbf{Our Model} & \textbf{WISARD \cite{wisard}} & \textbf{Forest PA \cite{ForestPA}} & \textbf{J48 Consolidated \cite{j48Consolidated}} & \textbf{LIBSVM \cite{libsvm}} & \textbf{FURIA \cite{furia}} & \textbf{Random Forest} & \textbf{REP Tree} & \textbf{MLP} & \textbf{Naive Bayes} & \textbf{Jrip} & \textbf{J48} \\ \hline
\textbf{TNR (BENIGN)}                  & \cellcolor{blue!25} 98.855\%           & 97.135\%                                             & 96.450\%                                                  & 93.355\%                                                                & 94.870\%                                             & 96.835\%                                           & 98.120\%                                                          & 95.165\%                                                & 92.650\%                                       & \cellcolor{red!25}66.545\%                                                    & 95.530\%                                         & 94.960\%                                       \\ \hline
\textbf{DR DDoS}                       & \cellcolor{blue!25}99.879\%           & \cellcolor{red!25}54.697\%                                             & 99.818\%                                                  & 93.212\%                                                                & 55.970\%                                             & 99.758\%                                           & 99.818\%                                                          & 99.788\%                                                & 91.212\%                                       & 93.879\%                                                    & 99.667\%                                         & 99.788\%                                       \\ \hline
\textbf{DR DoS slowloris}              & \cellcolor{blue!25}97.758\%           & 78.909\%                                             & 92.848\%                                                  & 95.030\%                                                                & \cellcolor{red!25}78.182\%                                             & 93.758\%                                           & 93.758\%                                                          & 92.727\%                                                & 78.485\%                                       & 82.667\%                                                    & 93.333\%                                         & 93.879\%                                       \\ \hline
\textbf{DR DoS Slowhttptest}           & \cellcolor{blue!25}93.841\%           &\cellcolor{red!25} 23.353\%                                             & 86.826\%                                                  & 83.832\%                                                                & 76.561\%                                             & 78.358\%                                           & 81.352\%                                                          & 75.364\%                                                & 88.537\%                                       & 70.060\%                                                    & 85.543\%                                         & 80.325\%                                       \\ \hline
\textbf{DR DoS Hulk}                   & 96.782\%           &\cellcolor{red!25} 67.600\%                                             & 93.945\%                                                  & 95.891\%                                                                & 73.709\%                                             & \cellcolor{blue!25}98.655\%                                           & 95.164\%                                                          & 92.218\%                                                & 86.891\%                                       & 73.782\%                                                    & 97.364\%                                         & 93.600\%                                       \\ \hline
\textbf{DR DoS GoldenEye}              & \cellcolor{blue!25}67.571\%           & \cellcolor{red!25}48.714\%                                             & 67.571\%                                                  & 67.143\%                                                                & 57.571\%                                             & 65.143\%                                           & \cellcolor{blue!25}67.571\%                                                          & 66.429\%                                                & 65.429\%                                       & 62.143\%                                                    & 63.857\%                                         & 67.286\%                                       \\ \hline
\textbf{DR Heartbleed}                 & \cellcolor{blue!25} 100\%          & 80.000\%                                             & \cellcolor{blue!25}100\%                                                 & 80.000\%                                                                &\cellcolor{red!25} 0.000\%                                              & 40.000\%                                           & \cellcolor{blue!25}100\%                                                         & \cellcolor{blue!25}100\%                                               & \cellcolor{red!25}0.000\%                                        & 80.000\%                                                    & 80.000\%                                         & 100\%                                      \\ \hline
\textbf{DR PortScan}                   & \cellcolor{blue!25}99.881\%           & 51.407\%                                             & 99.594\%                                                  & 99.046\%                                                                & \cellcolor{red!25}48.521\%                                             & 87.118\%                                           &\cellcolor{blue!25} 99.881\%                                                          &\cellcolor{blue!25} 99.881\%                                                & 48.521\%                                       & 99.499\%                                                    & 99.881\%                                         & 98.569\%                                       \\ \hline
\textbf{DR Bot}                        & 46.474\%           & 1.442\%                                              & 48.718\%                                                  &\cellcolor{blue!25} 52.083\%                                                                & \cellcolor{red!25}0.000\%                                              & 48.077\%                                           & 49.679\%                                                          & 47.756\%                                                & 51.282\%                                       & 29.968\%                                                    & 46.474\%                                         & 47.756\%                                       \\ \hline
\textbf{DR FTP-Patator}                & 99.636\%           & 0.000\%                                              & 99.727\%                                                  & \cellcolor{blue!25}100\%                                                               &\cellcolor{red!25} 0.000\%                                              & 99.636\%                                           & 99.727\%                                                          & 99.182\%                                                & 99.000\%                                       & 99.455\%                                                    & 99.545\%                                         & 99.545\%                                       \\ \hline
\textbf{DR SSH-Patator}                & 99.909\%           &\cellcolor{red!25} 0.000\%                                              &\cellcolor{blue!25} 100\%                                                 & 99.727\%                                                                &\cellcolor{red!25} 0.000\%                                              &\cellcolor{blue!25} 100\%                                          & 99.818\%                                                          &\cellcolor{blue!25} 100\%                                               & 99.727\%                                       & 99.182\%                                                    &\cellcolor{blue!25} 100\%                                        &\cellcolor{blue!25} 100\%                                      \\ \hline
\textbf{DR Web Attack - Brute Force}   & 73.265\%           &\cellcolor{red!25} 4.694\%                                              & 73.469\%                                                  & 55.102\%                                                                & 80.816\%                                             & 49.796\%                                           & 70.408\%                                                          & 70.816\%                                                &\cellcolor{blue!25} 90.408\%                                       & 5.102\%                                                     & 61.837\%                                         & 60.408\%                                       \\ \hline
\textbf{DR Web Attack - XSS}           & 30.625\%           & 1.250\%                                              & 34.375\%                                                  & 48.750\%                                                                & \cellcolor{red!25}0.000\%                                              & 38.750\%                                           & 37.500\%                                                          & 32.500\%                                                & 1.875\%                                        & \cellcolor{blue!25}91.875\%                                                    & 38.125\%                                         & 41.250\%                                       \\ \hline
\textbf{DR Web Attack - Sql Injection} & 50.000\%           & \cellcolor{red!25}0.000\%                                              & 50.000\%                                                  & \cellcolor{blue!25}100\%                                                               & \cellcolor{red!25}0.000\%                                              & 50.000\%                                           & \cellcolor{blue!25}100\%                                                         & 50.000\%                                                & 50.000\%                                       &\cellcolor{blue!25} 100\%                                                   & 75.000\%                                         & 50.000\%                                       \\ \hline
\textbf{DR Infiltration}               & \cellcolor{blue!25}100\%          & 50.000\%                                             & 83.333\%                                                  & 100\%                                                               & \cellcolor{red!25}0.000\%                                              & 83.333\%                                           & 83.333\%                                                          & 83.333\%                                                & 16.667\%                                       & 83.333\%                                                    & \cellcolor{blue!25}100\%                                        & 66.667\%                                       \\ \hline                                                       
\end{tabular}
\end{table*}
\vspace{-2pt}
\begin{table*}
\scriptsize
\centering
\caption{Overall Performance of our model and other classifiers}
\label{T_GP}
\begin{tabular}{|p{1.55cm}|p{0.8cm}|p{0.9cm}|p{0.9cm}|p{1cm}|p{0.85cm}|p{0.85cm}|p{1.15cm}|p{0.8cm}|p{0.85cm}|p{1cm}|p{0.75cm}|p{0.7cm}|}
\hline
                                       & \textbf{Our Model} & \textbf{WISARD \cite{wisard}} & \textbf{Forest PA \cite{ForestPA}} & \textbf{J48 Consolidated \cite{j48Consolidated}} & \textbf{LIBSVM \cite{libsvm}} & \textbf{FURIA \cite{furia}} & \textbf{Random Forest} & \textbf{REP Tree} & \textbf{MLP} & \textbf{Naive Bayes} & \textbf{Jrip} & \textbf{J48} \\ \hline

FAR      & \cellcolor{blue!25}1.145\%   & 2.865\%      & 3.550\%          & 6.645\%                 & 5.130\%      & 3.165\%      & 1.880\%              & 4.835\%         & 7.350\%    &  \cellcolor{red!25} 33.455\%         & 4.470\%     & 5.040\%    \\ \hline
DR (Overall)  & \cellcolor{blue!25}94.475\%  &  \cellcolor{red!25} 48.175\%     & 92.920\%         & 92.020\%                & 54.595\%     & 90.500\%     & 93.050\%             & 91.640\%        & 77.830\%   & 82.510\%         & 93.400\%    & 91.990\%   \\ \hline
Accuracy & \cellcolor{blue!25}96.665\%  &  \cellcolor{red!25} 72.655\%     & 94.685\%         & 92.688\%                & 74.733\%     & 93.668\%     & 95.585\%             & 93.403\%        & 85.240\%   & 74.528\%         & 94.465\%    & 93.475\%   \\ \hline

\textbf{Training Time} & 159.5 s & 13.47 s    & 110.64 s          & 105.46 s & 318.6 s  & 234.98 s       & 20.03 s   & 2.73 s & \cellcolor{red!25} 942.98 s   & \cellcolor{blue!25} 0.45 s  & 76.65 s & 8.34 s \\ \hline
\textbf{Test Time }    & 2.27 s  & 243.28 s   & 0.99 s            & 0.61 s   & \cellcolor{red!25} 343.96 s & 0.96 s         & 1.7 s     & 0.52 s & 1.61 s     & 12.39 s & \cellcolor{blue!25} 0.51 s  & 1.12 s \\ \hline 
\end{tabular}
\end{table*}
\vspace{-2pt}
Finally, each value $x_i$ of the feature j is normalized based on the following equation:

\begin{equation} \overline{x_i(j)} =\frac {x_i (j)-min⁡(x(j))}{max⁡(x(j))-min⁡(x(j))} \end{equation}

\subsection{Performance metrics}\label{sec_metrics}
IDS performance is evaluated based on its capability of classifying  network traffic into a correct type.
Table \ref{TCM}, also known as confusion matrix, shows all the  possible cases of classification.

To evaluate our model, we used two groups of metrics. The first group includes specific metrics: the detection rate (DR) of each type of attack and the true negative rate. The second one includes  global metrics: global detection rate, false alarm rate (FAR) and accuracy. The following equations summarize how to calculate theses metrics.

\small
\begin{equation} 
DR_{Attack Type}= \frac{TP_{Attack Type}}{TP_{Attack Type}+FN_{Attack Type}} 
\end{equation}
\vspace{-2pt}
\begin{equation} 
TNR_{BENIGN}= \frac{TN_{BENIGN}}{TN_{BENIGN}+FP_{BENIGN}} 
\end{equation}
\vspace{-2pt}
\begin{equation} FAR=  \frac {FP_{BENIGN}} {TN_{BENIGN}+ FP_{BENIGN} } \end{equation}
\vspace{-2pt}
\begin{equation} 
DR_{Overall}= \frac{TP_{Of-Each-Attack-Type}}{TP_{Of-Each-Attack-Type}+FN_{Of-Each-Attack-Type}} 
\end{equation}
\vspace{-2pt}
\begin{multline}
Accuracy= \frac{TP_{Of-Each-Attack-Type}+}{TP_{Of-Each-Attack-Type}+FN_{Of-Each-Attack-Type}+} \\
\frac{TN_{BENIGN}}{TN_{BENIGN}+FP_{BENIGN}}
\end{multline}
\normalsize
\vspace{-2.5em}

\subsection{Practical structure of our model}

Our model demands a pre-processing of data, different labelling of rows according to the model that is used and creation of artificial features that are outputs of two of the classifiers used. For training the first classifier we need to label the training data set based on the attacks classification provided in table \ref{T2} as "Attack" or 'Bening'. For the second classifier, we label each attack as follows : DDoS, DoS slowloris, DoS Slowhttptest, DoS Hulk, DoS GoldenEye, Heartbleed as "DoS"; FTP-Patator, SSH-Patator as "Brute-Force"; Web Attack - Brute Force, Web Attack - XSS, Web Attack - Sql Injection as "Web Attack". 
Finally, in order to train the third classifier, we use the labling used for the second classifier we add the outputs of the trained classifier 1 and classifier 2 as new features.

The choice of the three classifiers that compose our hierarchical model is the most important and critical step. To choose the best composition, that gives optimal performance, we tested several compositions with different classifiers. The results for all those combinations are quite lengthy and are not represnted in this article.  Following this demanding procedure, we opt for the following configuration: classifier 1 is REP Tree \cite{reptree} ; classifier 2 is Jrip \cite{jrip}; classifier 3 is Forest PA \cite{ForestPA}. 

\subsection{Comparative Study}
To evaluate our proposed model, we compare it with some well known classifiers and some recent ones namely J48, Jrip, Naive Bayes, MLP, REP Tree, Random Forest, FURIA \cite{furia}, LIBSVM \cite{libsvm}, J48 Consolidated \cite{j48Consolidated}, Forest PA \cite{ForestPA}, WISARD \cite{wisard}. In this comparative study we use the different metrics detailed in sub section \ref{sec_metrics}, in addition to training and test time.
\\
Table \ref{T_SP} summarizes the performance of our model compared to the other classifiers for different attacks and Benign traffic. It shows that our hierarchical model gives the highest true negative rate (TNR) with 98.855\% and the higest detection rate (DR) for six attacks type namely DDoS with 99.879\%, DoS slowloris with 97.758\%, DoS Slowhttptest with 93.841\%, DoS GoldenEye with 67.571\%, Heartbleed 100\%, PortScan 99.881\%, Infiltration 100\%. Moreover our hierarchical model is very close to the highest detection rate for two type of attacks namely  FTP Patator with 99.636\% and SSH Patator
with 99.909\%. For the rest of attacks type, our model gives an average performance compared to the other models. Overall, our model is the model that performs best for most of the different attacks type and it never has the lowest performance for any type of attack. 

Table \ref{T_GP} summarizes the global performance of our model as compared to the other classifiers. As shown in table  \ref{T_GP} our model gives the highest overall detection rate (DR Overall) with 94.475\% , the highest accuracy with 96.665\% , and lowest false alarm rate (FAR) with 1.145\%. The training time of our model is 195.5 seconds and the test time is 2.27 seconds, which represents an acceptable training and test time for a hybrid  hierarchical model especially when compared to simple models such as MLP and SVM.

\section{CONCLUSIONS}\label{sec_concl}
In this paper, we proposed an hierarchical intrusion detection system based on the combination of three different classifiers namely, REP Tree, JRip algorithm and Forest PA. The proposed model consists of three classifiers, where the outputs of the two of them to be used as inputs for the third. 
The evaluation using a real traffic data set 'CICIDS2017' showed that our hierarchical model outperformed different well known and recent machine learning models, giving the highest TNR and highest DR for seven of the attacks that exist in it. In overall, our model gives the highest DR with 94.457\%, the highest accuracy with 96.665\%, and the lowest FAR with 1.145\% while on the same its low computational time makes it easily incorporable in a soft real time system.

\addtolength{\textheight}{-12cm}  
\bibliographystyle{IEEEtran}
\bibliography{main}
\end{document}